\documentclass[twocolumn,showpacs,preprintnumbers,amsmath,amssymb]{revtex4}

\usepackage{graphicx}
\usepackage {amsmath}
\usepackage{hyperref}

\begin{document}

\title{Vortex state in Bose-Fermi mixture with attraction
between bosons and fermions}

\author{N.M. Chtchelkatchev}
\affiliation{Institute for High Pressure Physics, Russian Academy
of Sciences, Troitsk 142190, Moscow Region, Russia}

\author{S.-T. Chui}
\affiliation{Bartol Research Institute, University of Delaware,
Newark, DE 19716}

\author{V.N. Ryzhov}
\affiliation{Institute for High Pressure Physics, Russian Academy
of Sciences, Troitsk 142190, Moscow Region, Russia}

\date{\today}

\begin{abstract}
Vortex states in the mixture of ultracold atomic
clouds of bosons and fermions are investigated using the effective
Hamiltonian for the Bose subsystem. A stability of the Bose system
in the case of attractive interaction between components is
studied in the framework of variational Bose wave function and
Thomas-Fermi approximation. It is shown that the critical number
of bosons increases in the presence of the vortex.
\end{abstract}

\pacs{03.75.Fi,67.57.Fg,67.90.+z}


\maketitle

\section{Introduction}

Since the first realization of Bose-Einstein condensation (BEC)
in ultracold atomic gas clouds \cite{[1],[2],[3]}, studies in
this direction have yielded unprecedented insight into the
quantum statistical properties of matter. Besides the studies
using the bosonic atoms, growing interest is focused on the
cooling of fermionic atoms to a temperature regime where quantum
effects dominate the properties of the gas
\cite{sci99,sci01,sci02}. This interest is mainly motivated by
the quest for the crossover between a BEC and
Bardeen-Cooper-Schrieffer (BCS) superfluid in ultarcold atomic
Fermi gasses \cite{nat03_1,nat03_2,sci03}.

Strong {\it s}-wave interactions that facilitate evaporative
cooling of bosons are absent among spin-polarized fermions due to
the exclusion Pauli principle. So the fermions are cooled to
degeneracy through the mediation of fermions in another spin state
\cite{sci99,sci02,nat03_1,nat03_2,sci03} or via a buffer gas of
bosons \cite{sci01,prl1,sci02bf} (sympathetic cooling). The Bose
gas, which can be cooled evaporatively, is used as a coolant, the
fermionic system being in thermal equilibrium with the cold Bose
gas through boson-fermion interaction in the region of overlapping
of the systems.

However, the physical properties of Bose-Fermi mixtures are
interesting in their own rights and are the subject of intensive
investigations including the analysis of ground state
properties, stability, effective Fermi-Fermi interaction
mediated by the bosons, and new quantum phases in an optical
lattices
\cite{inst2,inst3,ChRyz04_PRA,ChRyz04_JETPL,stoof2,prl2}.
Several successful attempts to trap and cool mixtures of bosons
and fermions were reported. Quantum degeneracy was first reached
with mixtures of bosonic $^7$Li and fermionic $^6$Li atoms
\cite{sci01,prl1}. Later, experiments to cool mixtures of
$^{23}$Na and $^6$Li \cite{NaLi}, as well as $^{87}$Rb and
$^{40}$K \cite{sci02bf}, to ultralow temperatures succeeded.

Although Bose-Einstein condensation and supefluidity are closely
connected, they do not occur together in all cases. For example,
in lower dimensions one can observe superfluid without BEC. In
principle, rotational properties of a Bose-Fermi mixture could
directly reveal superfluidity in such systems. Quantized
vortices in a rotating gas provide direct conclusive evidence
for superfluidity because they are a consequence of of the
existence of a macroscopic wave function that describes the
superfluid. Recently such experimental evidence was obtained for
the superfluidity in strongly interacting Fermi gas
\cite{Fermi_sup}.

In this article we study the instability and collapses of the
trapped boson-fermion mixture due to the boson-fermion
attractive interaction in the presence of the quantized
vortices, using the effective Hamiltonian for the Bose system
\cite{ChRyz04_PRA,ChRyz04_JETPL}. We analyze quantitatively
properties of the $^{87}$Rb and $^{40}$K mixture with an
attractive interaction between bosons and fermions. The
stability of this system without vortices was recently studied
\cite{ChRyz04_JETPL}, and good agreement with experiment by
Modugno and co-workers \cite{sci02bf} was found. As was shown in
the experiment \cite{sci02bf}, as the number of bosons is
increased there is an instability value $N_{Bc}$ at which a
discontinues leakage of the bosons and fermions occurs, and
collapse of boson and fermion clouds is observed. In this
article we estimated the instability boson number $N_{Bc}$ for
the collapse transition in the presence of the vortices as a
function of the fermion number and temperature.

\section{Effective Bose Hamiltonian}

First of all we briefly discuss the effective boson Hamiltonian
\cite{ChRyz04_PRA,ChRyz04_JETPL}.  Our starting point is the
functional-integral representation of the grand-canonical
partition function of the Bose-Fermi mixture. It has the form
\cite{stoof2,popov1,stoof1}:
\begin{eqnarray}
Z&=&\int
D[\phi^*]D[\phi]D[\psi^*]D[\psi]\exp\left\{-\frac{1}{\hbar}\left(
S_B(\phi^*,\phi)+\right.\right.\\
\nonumber
&+&\left.\left.S_F(\psi^*,\psi)+S_{int}(\phi^*,\phi,\psi^*,\psi)\right)\right\}.
\label{1}
\end{eqnarray}
and consists of an integration over a complex field
$\phi(\tau,{\bf r})$, which is periodic on the imaginary-time
interval $[0,\hbar\beta]$, and over the Grassmann field
$\psi(\tau,{\bf r})$, which is antiperiodic on this interval.
Therefore, $\phi(\tau,{\bf r})$ describes the Bose component of
the mixture, whereas $\psi(\tau,{\bf r})$ corresponds to the Fermi
component. The term describing the Bose gas has the form:
\begin{eqnarray}
S_B(\phi^*,\phi)&=&\int_0^{\hbar\beta}d\tau\int d{\bf
r}\left\{\phi^*(\tau,{\bf r})\left(\hbar \frac{\partial}{\partial
\tau}-\frac{\hbar^2\nabla^2}{2 m_B} + \right.\right.\nonumber\\
 &+&V_B({\bf r})-\left.\left.\mu_B\right)\phi(\tau,{\bf
r})+ \frac{g_B}{2}|\phi(\tau,{\bf r})|^4\right\}. \label{2}
\end{eqnarray}
Because the Pauli principle forbids $s$-wave scattering between
fermionic atoms in the same hyperfine state, the Fermi-gas term
can be written in the form:
\begin{eqnarray}
S_F(\psi^*,\psi)&=&\int_0^{\hbar\beta}d\tau\int d{\bf
r}\left\{\psi^*(\tau,{\bf r})\left(\hbar \frac{\partial}{\partial
\tau}-\frac{\hbar^2\nabla^2}{2 m_F}+ \right.\right.\nonumber\\
&+&\left.\left.V_F({\bf r}) -\mu_F\right)\psi(\tau,{\bf r})
\right\}. \label{3}
\end{eqnarray}
The term describing the interaction between the two components of
the Fermi-Bose mixture is:
\begin{equation}
S_{int}(\phi^*,\phi,\psi^*,\psi)=g_{BF}\int_0^{\hbar\beta}d\tau\int
d{\bf r} |\psi(\tau,{\bf r})|^2|\phi(\tau,{\bf r})|^2, \label{4}
\end{equation}
where $g_B=4\pi \hbar^2a_B/m_B$ and $g_{BF}=2\pi
\hbar^2a_{BF}/m_I$, $m_I=m_B m_F/(m_B+m_F)$, $m_B$ and $m_F$ are
the masses of bosonic and fermionic atoms respectively, $a_B$ and
$a_{BF}$ are the $s$ wave scattering lengths of boson-boson and
boson-fermion interactions.

Integral over Fermi fields is Gassian, we can calculate this
integral and obtain the partition function of the Fermi system
as a functional of Bose field $\phi(\tau, {\bf r})$. It may be
shown \cite{ChRyz04_PRA,ChRyz04_JETPL} that the effective
bosonic Hamiltonian may be written in the form:
\begin{eqnarray}
H_{eff}&=&\int d{\bf
r}\left\{\frac{\hbar^2}{2m_B}|\nabla\phi|^2+(V_B({\bf
r})-\mu_B)|\phi|^2 +\right.\nonumber
\\ &+&\left.\frac{g_B}{2}|\phi|^4+f_{eff}(|\phi|)\right\}. \label{21}
\end{eqnarray}
where
\begin{eqnarray}
f_{eff}&=&-\frac{3}{2}\kappa\beta^{-1}\int_0^\infty\sqrt{\epsilon}
d\epsilon\ln\left(1+e^{\beta
(\tilde{\mu}-\epsilon)}\right)=\nonumber\\
&=&-\kappa\int_0^\infty\frac{\epsilon^{3/2}d\epsilon}{1+e^{\beta(\epsilon-\tilde{\mu})}},
\label{20}
\end{eqnarray}
and $\epsilon=p^2/2m_F$, $\tilde{\mu}=\mu_F-V_F({\bf
r})-g_{BF}|\phi(\tau,{\bf r})|^2$ and
$\kappa=2^{1/2}m_F^{3/2}/(3\pi^2\hbar^3)$.

 The first three terms in (\ref{21}) have the conventional
Gross-Pitaevskii \cite{pit} form, and the last term is a result
of boson-fermion interaction. In low temperature limit
$\tilde{\mu}/(k_BT)\gg 1$ one can write $f_{eff}(|\phi|)$ in the
form:
\begin{equation}
f_{eff}(|\phi|)=-\frac{2}{5}\kappa\tilde{\mu}^{5/2}-\frac{\pi^2}{4}\kappa(k_BT)^2
\tilde{\mu}^{1/2}. \label{25}
\end{equation}

As usual, $\mu_F$ can be determined from the equation
\begin{equation}
N_F=\int d{\bf r}n_F({\bf r}). \label{33}
\end{equation}
where
\begin{equation}
n_F({\bf r})=
\frac{3}{2}\kappa\int_0^\infty\frac{\sqrt{\epsilon}d\epsilon}{1+e^{\beta
(\epsilon-\tilde{\mu})}}. \label{31}
\end{equation}
At low temperatures we have:
\begin{equation}
n_F({\bf
r})=\kappa\tilde{\mu}^{3/2}+\frac{\pi^2\kappa}{8\tilde{\mu}^{1/2}}(k_BT)^2.
\label{32}
\end{equation}

In general case the Bose and Fermi systems have different
temperature scales, and Eqs. (\ref{25}) and (\ref{32}) may be
useful for studying the temperature behavior of Bose system,
including the calculation of the critical temperature. For
example, the characteristic temperature for the Bose system - the
transition temperature for the ideal Bose gas - is \cite{pit}:
$k_BT_c^0=0.94\hbar\omega_B(\lambda N_B)^{1/3}$. The Fermi
temperature for a pure system is \cite{butts}
$k_BT_F=\hbar\omega_F(6\lambda N_F)^{1/3}$. Taking into account
that $\omega_F=\sqrt{m_B/m_F}\omega_B$, one can see that for
$m_B>m_F$ and approximately the same numbers of bosons and
fermions one can safely use Eqs. (\ref{25}) and (\ref{32}) to
describe the behavior of the Bose system.

Let us consider now the $^{87}$Rb and $^{40}$K mixture  with an
attractive interaction between bosons and fermions \cite{sci02bf}.
The parameters of the system are the following: $a_B=5.25\,\, nm,
\ a_{BF}=-21.7^{+4.3}_{-4.8}\,\, nm$. K and Rb atoms were prepared
in the doubly polarized states $|F=9/2, m_F=9/2>$ and $|2,2>$,
respectively. The magnetic potential had an elongated symmetry,
with harmonic oscillation frequencies for Rb atoms
$\omega_{B,r}=\omega_B=2\pi\times 215 Hz$ and $
\omega_{B,z}=\lambda\omega_B=2\pi\times 16.3 Hz,
\omega_F=\sqrt{m_B/m_F}\omega_B\approx 1.47 \omega_B$, so that
$m_B\omega_B^2/2=m_F\omega_F^2/2=V_0$.  The collapse was found for
the following critical numbers of bosons and fermions:
$N_{Bc}\approx 10^{5}; N_K\approx 2\times 10^{4}$.

At the zero temperature limit, expanding $f_{eff}(|\phi|)$ up to
the third order in $g_{BF}$ we obtain the effective Hamiltonian in
the form:
\begin{eqnarray}
H_{eff}&=&\int d{\bf
r}\left\{\frac{\hbar^2}{2m_B}|\nabla\phi|^2+(V_{eff}({\bf
r})-\mu_B)|\phi|^2 +\right. \nonumber\\
&+&\left.\frac{g_{eff}}{2}|\phi|^4+\frac{\kappa}{8\mu_F^{1/2}}g_{BF}^3|\phi|^6\right\}.
\label{he}
\end{eqnarray}
where
\begin{eqnarray}
V_{eff}({\bf r})&=&\left(1-\frac{3}{2}\kappa\mu_F^{1/2}
g_{BF}\right)\frac{1}{2}m_B\omega^2_B\left(\rho^2+\lambda^2
z^2\right),\label{p1}\\
g_{eff}&=&g_B-\frac{3}{2}\kappa\mu_F^{1/2}g_{BF}^2,\label{p2}
\end{eqnarray}
and  $\rho^2=x^2+y^2$.

In principle, one can study the properties of a Bose-Fermi mixture
with the help of $f_{eff}$ (\ref{25}) without any expansion.
However, the form of the Hamiltonian (\ref{he}) gives the
possibility to get a clear insight into the physics of the
influence of the Fermi system on the Bose one (see discussion
below). It may be easily verified that the expansion of the
function $f(x)=(1+x)^{5/2}$ (see Eq. (\ref{25})) up to the third
order in $x$ gives a reasonably good approximation for $f(x)$ even
for rather large values of $x$, in contrast with the higher order
expansions, so one can safely use Eq. (\ref{he}) as a starting
point for the investigation of the properties of the Bose
subsystem.

In derivation of Eqs. (\ref{he}-\ref{p2}) we also use the fact
that due to the Pauli principle (quantum pressure) the radius of
the Bose condensate is much less than the radius of the Fermi
cloud $R_F\approx \sqrt{\mu_F/V_0}$, so one can use an expansions
in powers of $V_F({\bf r})/\mu_F$.

From Eq. (\ref{p1}) one can see that the interaction with Fermi
gas leads to modification of the trapping potential. For the
attractive fermion-boson interaction the system should behave as
if it was confined in a magnetic trapping potential with larger
frequencies than the actual ones, in agreement with experiment
\cite{sci02bf}. Boson-fermion interaction also induces the
additional attraction between Bose atoms which does not depend on
the sign of $g_{BF}$.

The last term in $H_{eff}$ (\ref{he}) corresponds to the
three-particle \emph{elastic} collisions induced by the
boson-fermion interaction. In contrast with \emph{inelastic}
3-body collisions which result in the recombination and removing
particles from the system \cite{kagan2}, this term for $g_{BF}<0$
leads to increase of the gas density in the center of the trap in
order to lower the total energy. The positive zero point energy
and boson-boson repulsion energy (the first two terms in Eq.
(\ref{he})) stabilize the system. However, if the central density
grows too much, the kinetic energy and boson-boson repulsion are
no longer able to prevent the collapse of the gas. Likewise the
case of Bose condensate with attraction (see, for example,
\cite{pit,kagan2,kagan1}), the collapse is expected to occur when
the number of particles in the condensate exceeds the critical
value $N_{Bc}$.

\section{Variational approach}

The qualitative (and even quantitative) picture of the stability
of the system can be obtained in the framework of the variational
approach. In the system without vortex the critical number
$N_{Bc}$ can be calculated using the well-known ansatz for the
Bosonic wave function \cite{pit}:
\begin{equation}
\phi({\bf
r})=\left(\frac{N_B\lambda}{w^3a^3\pi^{3/2}}\right)^{1/2}
\exp\left(-\frac{\left(\rho^2+\lambda^2
z^2\right)}{2w^2a^2}\right), \label{38}
\end{equation}
where $w$ is a dimensionless variational parameter which fixes the
width of the condensate and $a=\sqrt{\hbar/m_B\omega_B}$.

In this case the variational energy $E_B$ has the form:
\begin{equation}
\frac{E_B}{N_B\hbar\omega_B}=\frac{2+\lambda}{4}\frac{1}{w^2}+b
w^2+\frac{c_1N_B}{w^3}+\frac{c_2N_B^2}{w^6},\label{var}
\end{equation}
\begin{eqnarray*}
b&=&\frac{3}{4}\left(1-\frac{3}{2}\kappa\mu_F^{1/2}g_{BF}\right),\\
c_1&=&\frac{1}{2}\left(g_B-\frac{3}{2}\kappa\mu_F^{1/2}g_{BF}^2\right)
\frac{\lambda}{(2\pi)^{3/2}\hbar\omega_Ba^3},\\
c_2&=&\frac{\kappa}{8\mu_F^{1/2}}g_{BF}^3\frac{\lambda^2}{3^{3/2}
\pi^3\hbar\omega_Ba^6}.
\end{eqnarray*}
This energy is plotted in Fig.1 as a function of $w$ for several
values of $N_B$. It is seen that when $N_B<N_{Bc}$ there is a
local minimum of $E_B$ which correspond to a metastable state of
the system. This minimum arises due the competition between the
positive first three terms in Eq. (\ref{var}) and negative fourth
term. The local minimum disappears when the number of bosons $N_B$
exceeds the critical value which can be calculated by requiring
that the first and second derivatives of $E_B$ vanish at the
critical point. In this case the behavior of $E_B$ is mainly
determined by the second and fourth terms in Eq. (\ref{var}). For
$N_K= 2\times 10^{4}$ and $a_{BF}=-19.44\, nm$ we obtain
$N_{Bc}\approx 9\times10^4$ in a good agreement with the
experiment \cite{sci02bf}. It is interesting to note that the
critical number of Bose atoms in Bose-Fermi mixture is about two
orders larger than the critical number for the condensate with a
purely attractive interaction. For example, in the experiments
with trapped $^7$Li \cite{[3]} it was found that the critical
number of bosons is about $1000$.

If a vortex with angular momentum $\hbar l$ is present in the
system, the variational wave function may be written in the form
\cite{FetSvid}:
\begin{gather}
\phi({\bf r})=\sqrt{\frac{\lambda N_B}{(\omega
a)^5\pi^{3/2}}}\rho\exp\left(-\frac{\rho^2+\lambda^2
z^2}{2\omega^2 a^2}\right)e^{i\varphi l}. \label{var_vor}
\end{gather}

In this case the variational energy has the form:
\begin{gather*}
\frac {E_b}{N_B\hbar\omega_B}=\frac{2+\lambda^2+2l^2} 4 \frac 1
{\omega^2}+B\omega^2+\frac {C_1N_B}{\omega^3}+\frac{C_2
N_B^2}{\omega^6},
\\
B=\frac{5}{3}b,
\\
C_1=\frac{1}{2}c_1,
\\
C_2=\frac{2}{9}c_2.
\end{gather*}

\begin{figure}
\includegraphics{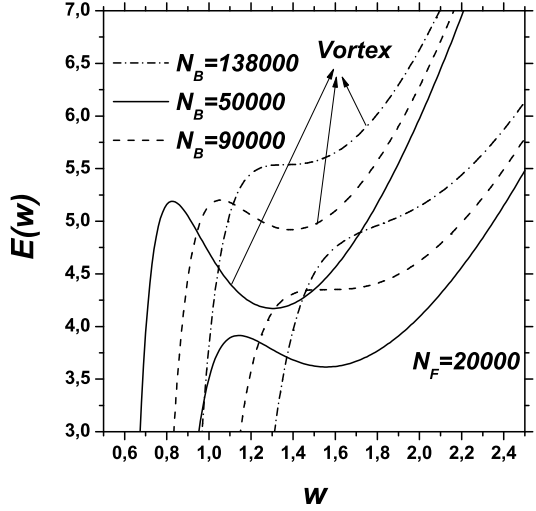}%
\caption{\label{fig:fig1}Variational energy
$\frac{E_B}{N_B\hbar\omega_B}$ as a function of $w$ for various
numbers of bosons. Upper lines correspond to the energy of the
system with a vortex.}
\end{figure}

From Fig. 1 one can see that the stability of the system with a
vortex is higher (in this case the critical number of bosons
$N_{Bc}\approx 135\times10^3$) due to the centrifugal effect
(compare the first terms in Eqs. (\ref{var}) and (\ref{var_vor})
and decreasing the number density of the bosons in the center of
the trap (fourth terms in Eqs. (\ref{var}) and (\ref{var_vor}).

In Eq. (\ref{var}) we use $\mu_F^0=\hbar\omega_F[6\lambda
N_F]^{1/3}$ as the chemical potential of the Fermi system $\mu_F$.
The corrections to $\mu_F$ due to interaction with the Bose system
have the form: $\mu_F=\mu_F^0[1+m_1]$, where $m_1=1/2\kappa
g_{BF}{\mu_F^0}^{1/2}N_B/N_F$.  It may be shown that $m_1\approx
0.09$ for the values of the parameters used in these calculations.
The correction to the chemical potential for the vortex state is
the same.

Upon increasing the number of fermions, the repulsion between
bosons decreases leading to the collapse for the smaller numbers
of the bosonic atoms. In Fig. 2 the critical number of bosons
$N_{Bc}$ is represented as a function of the number of fermions
for both cases: the state without vortex and the state with the
vortex.

\begin{figure}
\includegraphics{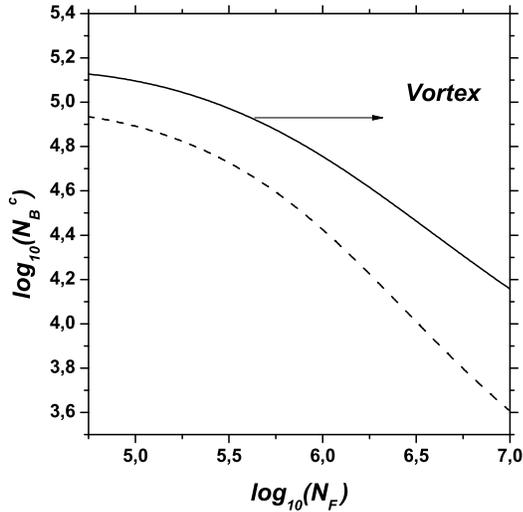}%
\caption{\label{fig:fig2}Critical number of bosons $N_{Bc}$ as a
function of the number of fermions $N_F$ at $T=0$.}
\end{figure}

It should be noted that the critical number of bosons $N_{Bc}$
is extremely sensitive to the precise value of the boson-fermion
$s$ wave scattering length. This is illustrated in Fig. 3 in
\cite{CRT_JETP}.

The critical velocity of the vortex state can be found from the
formula \cite{FetSvid}: $\Omega_c=(E_b(l=1)-E_b)/\hbar N_B$. In
Figs. 4 and 5 $\Omega_c$ is shown as a function of $N_B$ and
$N_F$.

\begin{figure}
\includegraphics{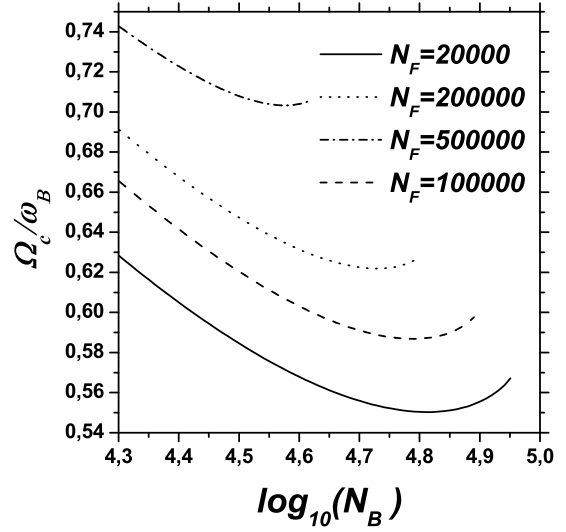}%
\caption{\label{fig:fig3}Critical angular vortex velocity as a
function of a logarithm of the number of bosons $N_B$.}
\end{figure}

\begin{figure}
\includegraphics{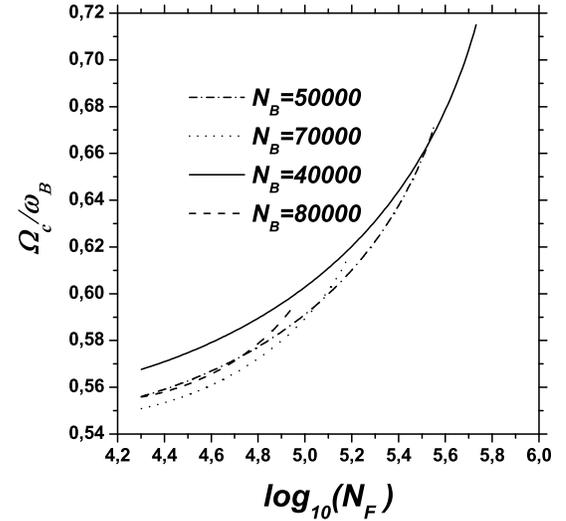}%
\caption{\label{fig:fig4}Critical angular vortex velocity as a
function of a logarithm of the number of fermions $N_F$.}
\end{figure}

The curves in Figs. 3 and 4 are ending at the critical points
which correspond to the critical numbers of bosons and fermions
at which the collapse of the system occurs. The minima in Fig. 4
and intersections of the curves in Fig. 5 seem to be artifacts
of variational approach to calculation of the critical angular
velocity in the vicinity of the critical points. It should be
noted that the number of bosons in the system is rather large.
In this case the qualitatively and even quantitatively correct
results may be obtained in the framework of the Thomas-Fermi
approximation (TFA).

\section{Thomas-Fermi approximation}

In the Thomas-Fermi approximation the kinetic energy terms are
ignored. It has been shown that in the case of one component
condensates the TFA results agree well  with the numerical
calculations for large particle numbers, except for a small
region near the boundary of the condensate
\cite{pit,FetSvid,BP}. In fact, even for a small number of
particles the TFA still usually gives qualitatively correct
results. The TFA provides an excellent starting point of study
of the vortex states in Bose condensates (see, for example,
\cite{sinha,CRT_JETP,CRT_PRA}).

The Gross-Pitaevski equation that follows from the effective
hamiltonian (\ref{he}):
\begin{gather}
\left(-\frac{\hbar^2}{2m_B}\triangle\phi+(V_{\rm
eff}-\mu_B)+g_{\rm eff}|\phi|^2+\right.\\
\left.+\frac{3\kappa}{8\mu_F^{1/2}}g_{BF}^3|\phi|^4\right)\phi=0.
\label{GPe}
\end{gather}

In TFA one can neglect the kinetic energy. In this case boson
density has the form:
\begin{gather}
|\phi|^2=\theta(\mu_B-V_{\rm eff})\frac{g_{\rm
eff}\left(-1+\sqrt{1+(\mu_B-V_{\rm eff})\frac{3\kappa
g_{BF}^3}{g_{\rm eff}^2 2\mu_F^{1/2}}}\right)}{\frac{3\kappa
g_{BF}^3}{4\mu_F^{1/2}}}\label{den1}
\end{gather}

In the limit $g_{BF}\to 0$ one has the conventional expression
for the boson density \cite{pit,FetSvid}:
\begin{gather*}
|\phi|^2=\theta(\mu_B-V_{\rm eff})\frac {(\mu_B-V_{\rm eff})}
{g_{\rm eff}}.
\end{gather*}

Equation (\ref{den1}) for the boson density $n({\bf
r})=|\phi({\bf r})|^2$ may be rewritten in the form:
\begin{gather}
n({\bf
r})=n(0)\left(1-\sqrt{1+\frac{\bar{x}^2+\bar{y}^2+\lambda^2\bar{z}^2-R^2}
{R_{max}^2}}\right).\label{den2}
\end{gather}

Here
\begin{gather*}
\bar{r_i}=r_i/a_h; \,\, R^2=\frac{2\mu_B}{c_0m_B\omega_B^2
a_h^2}; \,\,
\\ n(0)=-\frac{4}{3}\frac{g_{eff}\mu_F^{1/2}}{\kappa g_{BF}^3};
\,\, R_{max}^2=-\frac{4}{3}\frac{g_{eff}^2\mu_F^{1/2}}{\kappa
g_{BF}^3c_0m_B\omega_B^2 a_h^2}; \\
c_0=\left(1-\frac{3}{2}\kappa\mu_F^{1/2} g_{BF}\right).
\end{gather*}

$R_{max}$ is the radius of the condensate. The critical number
of bosons can be determined from the condition that the
expression under the square root in (\ref{den2}) is positive. In
Fig. 5 the critical number of bosons is shown as a function of
the number fermions (lower curve).

\begin{figure}
\includegraphics{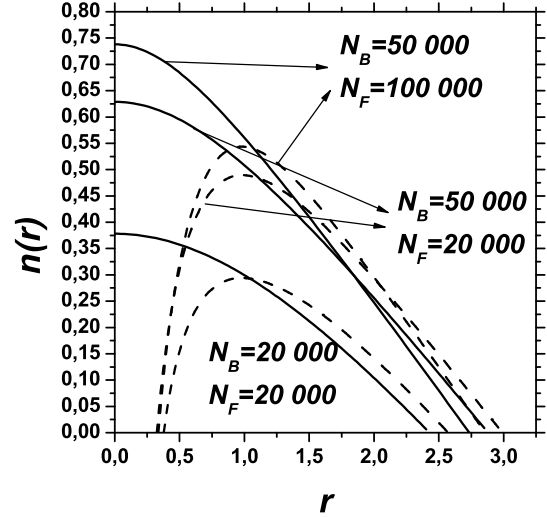}%
\caption{\label{fig:fig5}Critical number of Bosons $N_{Bc}$ as a
function of the number of Fermions $N_F$ in TFA.}
\end{figure}

Let us now consider a trap rotating with frequency $\Omega$
along the $z$-axis. For vortex excitation with angular momentum
$\hbar l$ , the condensate wave function is given by
\begin{equation}
\phi_{l}({\bf r})=\sqrt{n_l({\bf r})}e^{i l\varphi}.
\label{wf_rot}
\end{equation}
In a rotating frame the energy functional of the system is
\begin{equation}
E_{rot}(l)=E(\phi_l)+\int\,d^3 r (\phi_{l}^*)i\hbar\Omega
\partial_\varphi (\phi_{l}). \label{e_rot}
\end{equation}

After substituting the wave function for the vortex excitation
(\ref{wf_rot}) in Eq. (\ref{e_rot}), the effective confinement
potential for the bosons becomes $l^2\hbar^2/2m\rho^2+ V$, where
$V=m\omega(\rho^2+ \lambda^2 z^2)/2$ and $\rho^2=x^2+y^2$. So
within the TFA the density of the vortex state has the form:
\begin{gather}
n_l({\bf
r})=n(0)\left(1-\sqrt{1+\frac{\bar{\rho}^2+\lambda^2\bar{z}^2+
\frac{l^2}{c_0\bar{\rho}^2}-R^2} {R_{max}^2}}\right).
\label{den3}
\end{gather}
 The
important new qualitative feature of a vortex in the TFA is the
appearance of a small hole of radius $\xi$, $\xi\propto
l/R_{max}$, but the remainder of the condensate density is
essentially unchanged. The fractional change in the chemical
potentials caused by the vortex $(\mu'(l)-\mu')/\mu'$ can be
shown to be small \cite{sinha,FetSvid}, of the order of
$1/N^{4/5}$. In the calculation of physical quantities involving
the condensate density it is sufficient to retain the no-vortex
density and simply cut off any divergent radial integrals at the
appropriate core sizes $\xi=l/R_{max}$.  Note that using the
unperturbed density for calculation of the vortex properties
corresponds to the hydrodynamic limit.

In Fig. 6 the density distribution of bosons calculated from
Eqs. (\ref{den2}) and (\ref{den3}) is shown as a function of
radius for different numbers of bosons and fermions.

\begin{figure}
\includegraphics{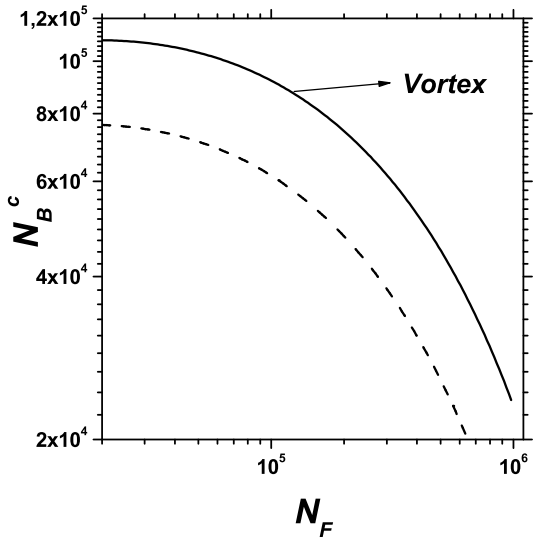}%
\caption{\label{fig:fig6}Density distribution of bosons as a
function of radius for different numbers of bosons and fermions
calculated in TFA.}
\end{figure}

The critical number of bosons in the presence of a vortex can be
calculated from the condition that the expression under the square
root in (\ref{den3}) is positive. This number is shown in Fig. 5
(the upper curve) as a function of the number of fermions. As was
expected the presence of the vortex increases the stability of the
system against the collapse transition.

\begin{figure}
\includegraphics{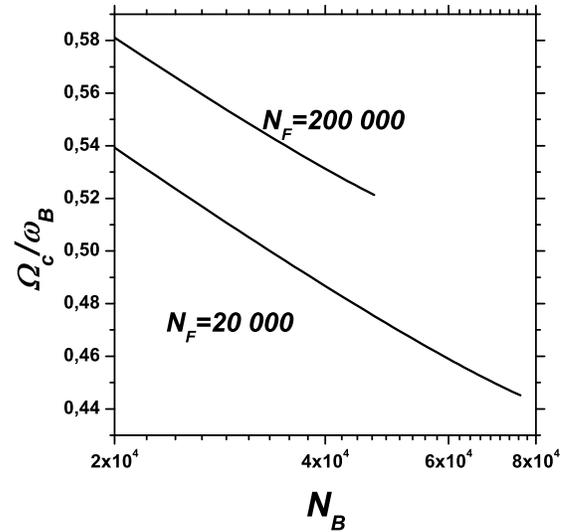}%
\caption{\label{fig:fig7} Critical angular vortex velocity as a
function of a logarithm of the number of bosons $N_B$ in TFA.}
\end{figure}

From Eqs. (\ref{e_rot}-\ref{wf_rot}) and (\ref{he}-\ref{den2})
and Eq. $\Omega_c=(E_b(l=1)-E_b)/\hbar N_B$ one can find the
critical angular velocity in TFA. The critical angular velocity
as a function of the numbers of bosons and fermions
correspondingly is shown in Figs. 7 and 8. One can see that the
results are free from the mistakes which are present in the
variational approach (see Figs. 3 and 4).

\begin{figure}
\includegraphics{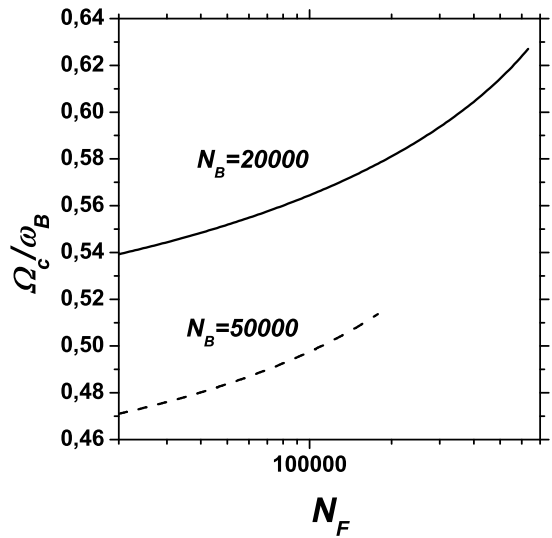}%
\caption{\label{fig:fig8} Critical angular vortex velocity as a
function of a logarithm of the number of fermions $N_F$ in TFA.}
\end{figure}

\section{Final remarks}
 Finally, we make a short remark on the nature of the
collapse transition. In this article we found the instability
point of the Bose-Fermi mixture with attractive interaction
between components with and without vortex. A strong rise of
density of bosons and fermions (see Eq. (\ref{32})) in the
collapsing condensate enhances intrinsic inelastic processes, in
particular, the recombination in 3-body interatomic collisions,
as is the case for the well-known $^7$Li condensates
\cite{kagan2}. In the presence of a vortex there appears a hole
in the middle of the condensate. This reduces the maximum
density of the condensate (see Fig. 6) and increases the
critical number of bosons. Recently M. Yu. Kagan and coworkers
suggested the new microscopic mechanism of removing atoms from
the system which is specific for the Bose-Fermi mixtures with
attraction between components and is based on the formation of
the boson-fermion bound states \cite{MKag}. It seems that the
description of the evolution of the collapsing condensate should
include both these mechanisms.

\section{Acknowledgement}
VNR acknowledges valuable discussions with M. Yu. Kagan. The work
was supported in part by the Russian Foundation for Basic Research
(Grant No 05-02-17280) and NWO-RFBR Grant No 047.016.001. STC is
partly supported by a grant from NASA. NMC acknowledges the
support from RFBR (Grant No 04-02-08159), Government Grant
ÌÊ-4611.2004.2 and Russian Science Support Foundation.

\end{document}